# Inverse Designed THz Spectral Splitters

Sourangsu Banerji, Yu Shi, Vivian Song-En Su, Udayan Ghosh, Jacqueline Cooke, *Student Member, IEEE*, Yong Lin Kong, *Member, IEEE,* Lei Liu, and Berardi Sensale-Rodriguez, *Senior Member, IEEE*

*Abstract*— This letter reports proof-of-principle demonstration of 3D printable, low-cost, and compact THz spectral splitters based on diffractive optical elements (DOEs) designed to disperse the incident collimated broadband THz radiation (0.5 THz - 0.7 THz) at a pre-specified distance. Via inverse design, we show that it is possible to design such a diffractive optic, which can split the broadband incident spectrum in any desired fashion, as is evidenced from both FDTD simulations and measured intensity profiles using a 500-750 GHz VNA. Due to its straightforward construction without the usage of any movable parts, our approach, in principle, can have various applications such as in portable, low-cost spectroscopy as well as in wireless THz communication systems as a THz demultiplexer.

## I. Introduction

Spectrometers are widely used in optical characterization, biomedical diagnostics, astronomical exploration, chemical analysis, etc. [1, 2]. Traditional spectrometers harness gratings (typically one for each blazing wavelength or multi-orders) or prisms, coupled with other mechanical and electronic parts, to disperse and detect the incident spectrum [1]. The necessity to assemble multiple components within a single system renders them bulky, sensitive to alignment, and suffering from low throughput due to cross talk. Hence, they are not the optimal choices for broad commercial and industrial applications, where small footprint and simple hardware are required [2]. From such a perspective, coded apertures have been previously developed to construct compact computational spectrometers over both visible [3, 4] and infrared [5] regimes. However, such an approach is limited to narrow spectral bands and sensitive to noise.

Nonetheless, from a computational perspective, we show that a single broadband diffraction grating can be inverse designed via numerical optimization of the grating's surface topology to split the incident broadband radiation under arbitrary splitting conditions. This inverse design formulation is a very common technique employed in earlier works by Lohmann [6], Wai-Hon Lee [7], Lesem, Hirsch, and Jordan [8] as well as extensively used in digital holography by Bryngdahl and others [9, 10], where a hologram is designed to project a pre-defined intensity pattern and allow the phase in the image plane to be arbitrary, i.e. "*phase as a free parameter*" concept [11].

In some of the preliminary work, we have numerically shown that designing spectral splitters to split incident broadband THz radiation spatially is possible [12, 13]. In this letter, we demonstrate the experimental operation of four such splitter structures with careful geometric optimization of the heights of each individual element ("pixel") of a diffraction phase grating for a broadband THz spectrum, i.e., from 0.5 to 0.7 THz.

## II. Methods

### A. Theory

From a fundamental standpoint, the idea of designing such a splitter structure (Fig. 1(a)) is related to inverse scattering, which forms the basis for the inverse design methodology, as discussed in the latter part of the previous section. Indeed, if one considers two parallel planes: a "grating" and an "observation" plane. The goal is to design a phase grating structure that produces the desired "spectral split" field pattern when illuminated by broadband radiation. A phase grating structure can now simply be obtained using backpropagation. However, this is an ill-posed problem since there can be many grating (phase transmittance function of a diffraction grating) structures that approximately give the same observation field pattern (e.g., by adding to the phase grating's spatial modes corresponding to evanescent waves).

Furthermore, the finite extent of the THz detector in the observation plane can also cause ambiguities in the grating's structural topology that depend nonlinearly on the position on the grating plane [14, 15]. This categorically points to the fact that the choice of an ideal phase transmittance function for the grating is not necessarily unique. As a result, we argue that optimization is better suited to choosing an appropriate surface topography, as is the case for inverse design problems.

### B. Design

The dimension of the splitter is taken to be 40 mm by 40 mm in length and width respectively (since the THz beam size (Gaussian beam radius x 2) to be used for experimentation was calculated to be ~32 mm); consisting of multilevel pixels having maximum thickness $h_{max}$= 2000 μm, minimum thickness

This work was primarily supported by the NSF awards: ECCS # 1936729 and MRI #1828480. This work was also partially supported by the AFOSR award FA9550-18-1-033

S. Banerji, V. S-E. Su, U. Ghosh, Shu, J. Cooke, Y. L. Kong, and B. Sensale-Rodriguez are with the University of Utah, Salt Lake City, UT 84108 USA (e-mail: berardi.sensale@utah.edu).

Y. Shi and L. Liu are with the Electrical Engineering Department, University of Notre Dame, Notre Dame, IN 46556 USA.

$h_{min}$ = 200 μm, and height level step Δh = 200 μm; which sets the number of distinct height levels (P) to P = 10. The pixels have a width w = 400 μm; this sets the number of pixels, i.e., N = 200. The THz spectral splitters were designed to split the incident broadband radiation at a distance d = [35 mm, 50 mm]. In total, four different spectral splitters were designed to portray the robust and dynamic splitting capability of our inverse designed based formulation. Fig. 1(b-e) depicts the pixel height profile along with relevant geometric dimensions for all the designs.

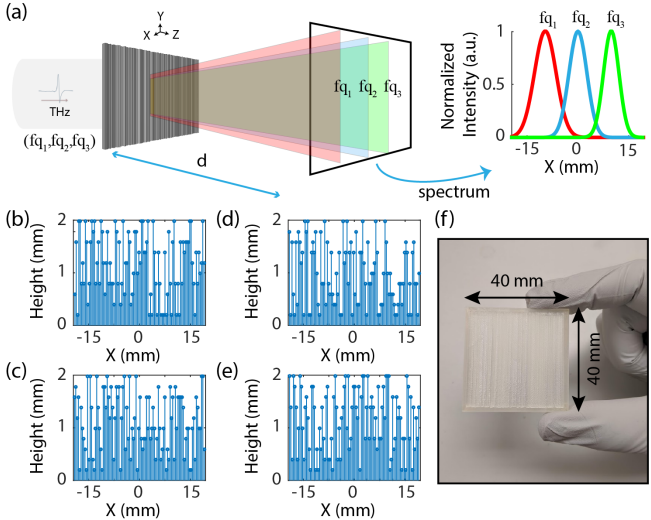

Fig. 1. (a) Schematic of the spectral splitter with a splitting distance d = [35 cm, 50 cm] under broadband illumination ($\lambda_1$ = 0.6 mm [0.5 THz], $\lambda_2$ = 0.5 mm [0.6 THz], and $\lambda_3$ = 0.4 mm [0.7 THz]). The designed structure splits the incoming THz wave into a series of spatially separated lines at the pre-specified design distance. (b-e) Pixel height distribution for the spectral splitter with a maximum pixel height of 2 mm. The dimensions (length and width) of the spectral splitter was 40 mm. (f) Optical micrograph of a fabricated sample.

Speaking of optimization parameters, the target function was defined via dictating the spatial location of each specific frequency at the observation plane. Detailed specifics about the Figure of Merit (FoM) and target function are provided in [12]. PLA was the material of choice when defining the optical constants in the design. A value of n = 1.357 and k = 0.051 at 0.4 THz, n = 1.360 and k = 0.067 at 0.5 THz and n = 1.367 and k = 0.097 at 0.6 THz were employed [16].

### C. Simulation

The optimization and simulation of the structures were carried out using both MATLAB and Lumerical FDTD solutions. The optimization routine in MATLAB generates the corresponding Lumerical FDTD script file during each iteration, which, in turn, runs the FDTD simulation. After the simulation gets completed, data from Lumerical FDTD is loaded back into MATLAB, which evaluates the FoM and terminates only if a suitable convergent solution is reached. The details of the optimization algorithm are provided in [12]. For FDTD, a gaussian illumination is considered to impinge upon the structure. PML boundary conditions are considered in all directions. No symmetry condition could be imposed to speed up the computation due to the inherent asymmetric nature of the problem. However, a more convenient 2D FDTD is carried out instead of a full 3D FDTD simulation since the structure is 1D in the x-direction with no variation across y-direction. The mesh accuracy was to ~λ/35 in the 2D FDTD simulation.

### D. Fabrication

A 3D computer-aided design (CAD) model of THz spectral splitter (40 mm x 40 mm) was first created with Solidworks (Dassault Systèmes). Optimized pixel height was considered for modeling the grating of the splitter. Standard Tessellation Language (STL) files were then digitally sliced and converted into G-code for generating the printing path using Ultimaker Cura. The converted and optimized G-code were printed with a multi-material fused deposition modeling (FDM) 3D Printer (Ultimaker 3) with Polylactic Acid (PLA) filament. The optimum structure is determined, and a 100% infill density and a layer height of 0.06 mm were used for the results. An optical micrograph of one such 3D printed sample is shown in Fig. 1(f).

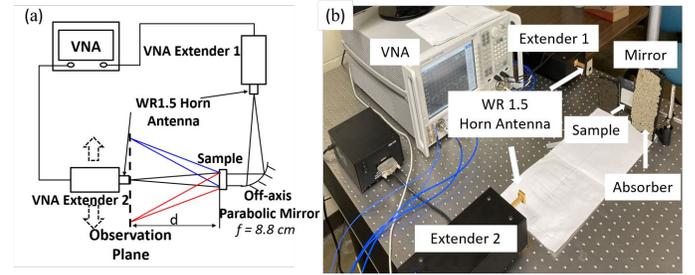

Fig. 2. (a) Schematic and (b)the experimental setup used to characterize the THz spectral splitters.

### E. Measurement

The spectral splitter was experimentally characterized in a quasi-optical THz measurement system (as shown in Fig.2), including an Agilent N5245A Vector Network Analyzer (VNA) with two WR-1.5 bands (500–750 GHz) frequency extenders. For the purpose of this measurement, the incident electromagnetic wave (EM) was generated by port 1 of the VNA and provided through a WR-1.5 diagonal horn antenna. The THz beam, which was collimated by an off-axis parabolic mirror (focal length f = 8.8 cm), can then be approximately treated as a plane wave. The beam waist was measured to be 31.5 mm (radius or diameter) after the parabolic mirror so that the incoming THz beam can illuminate the entire sample area (40 mm by 40 mm). An absorber was placed around the sample to block the surrounding EM wave to minimize the interference. The S-parameters were then measured along the observation plane at a distance "d" from the sample and the corresponding transmitted power was collected. For samples 1 and 2, the transmitted power spectrum was obtained at d = 35 cm, whereas d becomes 50 cm for samples 3 and 4 measurements.

## III. RESULTS AND DISCUSSION

Fig. 3(a-b) and Fig. 4(a-b) respectively depict the simulated and measured spectral maps of two spectral splitters which were designed to split incident THz frequencies in a regular sequence (gradual split) across the observation plane at a pre-determined distance of 35 cm and 50 cm for the designed frequencies of 0.5 THz, 0.6 THz, and 0.7 THz. In addition to this, Fig. 3(c-d) and Fig. 4(c-d) portrays the simulated and measured spectral maps of a separate set of two more spectral splitter designs which were designed to split the same incident THz radiation in an arbitrary sequence (random split) across the observation plane at a pre-determined distance of 35 cm (Fig. 3(c-d)) and 50 cm (Fig. 4(c-d)) subsequently. In general, from the spectral plots of both Fig. 3 and Fig. 4, it is evident that the split does happen even though the exact spatial location might be a bit off in the measured spectral profile.

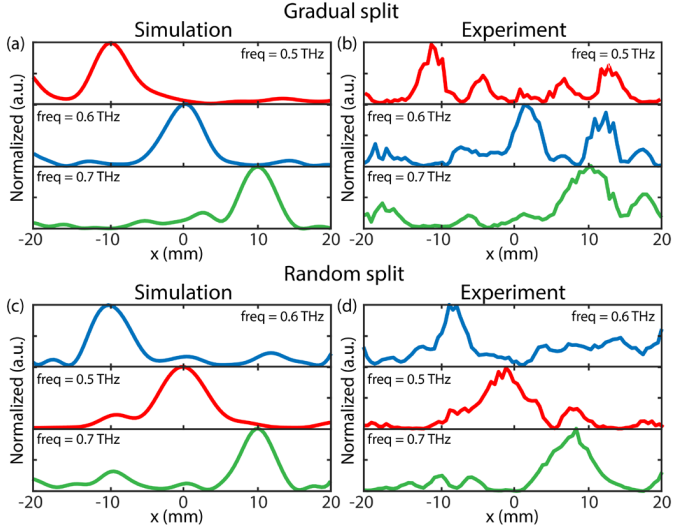

Fig. 3. Spectral splitter designs for a gradual split (sample 1 and sample 2) with (a) simulated and (b) measured spectral map at a splitting distance d = 35cm under broadband illumination ($\lambda_1$ = 0.6 mm [0.5 THz], $\lambda_2$ = 0.5 mm [0.6 THz], and $\lambda_3$ = 0.4 mm [0.7 THz]). The corresponding spectral splitter designs for a random split (c) simulated and (d) measured spectral map at a splitting distance d = 35 cm under broadband illumination ($\lambda_1$ = 0.5 mm [0.6 THz], $\lambda_2$ = 0.6 mm [0.5 THz], and $\lambda_3$ = 0.4 mm [0.7 THz]).

This can be attributed to the fact that, under alike geometric conditions, the spectral resolution offered by the non-regular-sequence design tends to be higher than that by the design with regular spectral split due to the spectral correlation function. In principle, the spectral correlation function measures how similar the diffraction patterns are at two distinct wavelengths. For regular-sequence spectral splitter designs, the spatial-spectral map changes relatively smoothly in contrast to that of a random design, which often experiences abrupt variations. Therefore, the correlation function for random designs becomes narrower, which makes it easier to distinguish between frequencies. These observations are consistent with those of earlier works reported in the literature for such diffractive optic-based splitter designed at optical frequencies [17-19].

However, here an important observation can be made. The splitter designs, which had a random non-monotonic split of the incident THz frequencies, depicted relatively better performance than their regular-sequence counterparts. In general, sample 2 and sample 4 have a clean spectral map with suppressed side lobes with respect to both sample 1 and sample 3. In fact, one can observe that sample 2 has the best split performance amongst all the four designs.

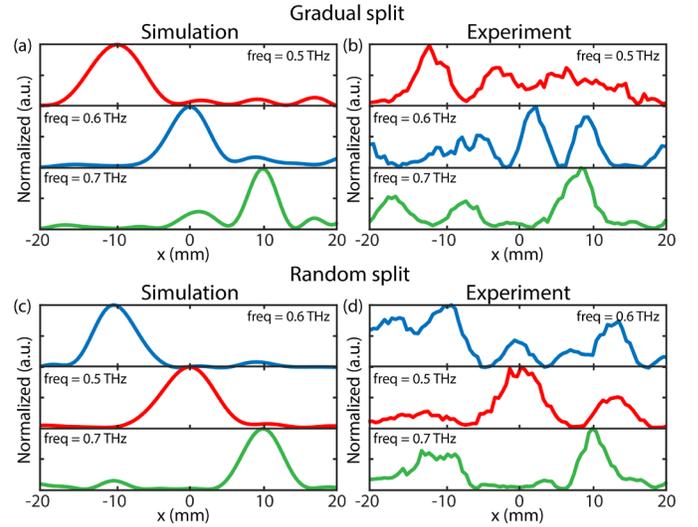

Fig. 4. Spectral splitter designs for a gradual split (Sample 3 and Sample 4) with (a) simulated and (b) measured spectral map at a splitting distance d = 50 cm under broadband illumination ($\lambda_1$ = 0.6 mm [0.5 THz], $\lambda_2$ = 0.5 mm [0.6 THz], and $\lambda_3$ = 0.4 mm [0.7 THz]). The corresponding spectral splitter designs for a random split (c) simulated and (d) measured spectral map at a splitting distance d = 50 cm under broadband illumination ($\lambda_1$ = 0.5 mm [0.6 THz], $\lambda_2$ = 0.6 mm [0.5 THz], and $\lambda_3$ = 0.4 mm [0.7 THz]).

Finally, the appearance of substantial side lobes in the measured results can be attributed to two reasons: (a) an imperfection in fabrication due to the inherent limitation of fused deposition modeling 3D printing (typical maximum layer resolution [z] of ~ 20 µm and planar resolution [x, y] ~ 400 µm); and (b) an imperfection in the measurement setup. A theoretical statistical study was already conducted on a very similar structure, i.e., a multilevel diffractive lens in [16] to showcase how the performance of the multilevel structure degrades with the inherent imperfection of fabrication, i.e., error in height and width of each pixel as well as density variations leading to index non-uniformities; and hence a detailed discussion is omitted here. From the measurement perspective, the receiver VNA extender was manually moved along the desired plane which will introduce additional misalignment errors. Besides, the receiver WR1.5 horn antenna is not a perfect point detector. Instead, its radiation pattern is also a Gaussian beam with certain beam width and substantial side lobes, which may broaden the main lobe and introduce side lobes in the measurements. The insertion loss of the splitter is estimated to be 12 dB at the spectral peak position, which is mainly from the dielectric loss of the material with a complex refractive index.

## CONCLUSION

In conclusion, we have demonstrated compact THz spectral splitters via inverse design which is capable of splitting incident broadband THz waves in free space with appreciable accuracy. Barring the challenges associated with the current state-of-the-art 3D printing technology and the usage of better measurement facilities, this simple proof-of-concept demonstration of such spectral splitters evidences the fact that the proposed structures can be crucial in enabling portable, low-cost spectrometers as well as in wireless communications as THz demultiplexers.

## REFERENCES

[1] Bacon, C. P., et. al. "Miniature spectroscopic instrumentation: applications to biology and chemistry," Review of Scientific Instruments 75, (2004): 1–16.
[2] Khorasaninejad, M., et. al., "Super-dispersive off-axis meta-lenses for compact high resolution spectroscopy." Nano letters 16.6 (2016): 3732-3737.
[3] Gehm, M. E., et. al., "Static two-dimensional aperture coding for multimodal, multiplex spectroscopy," Appl. Opt. 45(13), 2965–2974 (2006).
[4] Feller, S. D., et. al., "Multiple order coded aperture spectrometer," Opt. Express 15(9), 5625–5630 (2007).
[5] Fernandez, C., et. al., "Longwave infrared (LWIR) coded aperture dispersive spectrometer," Opt. Express 15(9), 5742–5753 (2007).
[6] Lohmann, A. W., et. al., "Optical information processing" (Ilmenau University, 1978).
[7] Lee, W.H. "Diffractive optics for data storage," Proc. SPIE 2383, 390–395 (1995).
[8] Lesem, L. B., et. al., "The kinoform: a new wavefront reconstruction device," IBM J. Res. Dev. 13(2), 150–155 (1969).
[9] Wyrowski, F., et. al., "Digital holography as part of diffractive optics," Rep. Prog. Phys. 54(12), 1481–1571 (1991).
[10] Soifer, V. A., "Computer design of diffractive optics" (Elsevier, 2012).
[11] Banerji, S., et. al., "Imaging over an unlimited bandwidth with a single diffractive surface," arXiv: 1907.06251 [physics.optics] (2019)
[12] Banerji, S., et. al., "A Computational Design Framework for Efficient, Fabrication Error-Tolerant, Planar THz Diffractive Optical Elements" Scientific Reports. 9, (2019): 5801.
[13] Banerji, S., et. al., "3D-printed diffractive terahertz optical elements through computational design." Proc. SPIE 10982, 109822X (2019)
[14] Banerji, S., et. al., "Imaging with flat optics: metalenses or diffractive lenses?" Optica 6, 805-810 (2019).
[15] Maretske, S., "Locality estimates for Fresnel-wave-propagation and instability of x-ray phase contrast imaging with finite detectors," arXiv:1805.06185 (2018).
[16] Banerji, S., et. al., "Impact of fabrication errors and refractive index on multilevel diffractive lens performance." Scientific reports 10, no. 1, 1-8 (2020).
[17] Wang, P., et. al., "Computational spectrometer based on a broadband diffractive optic." Opt. Exp. 22, 14575–14587 (2014).
[18] Yue, J., et. al. "High-throughput deconvolution-resolved computational spectrometer." Chi. Opt. Lett. 12, 043001 (2014).
[19] Wang, P., et. al., "Computational spectroscopy via singular-value decomposition and regularization." Opt. Exp. 22, 21541–21550 (2014).